\documentclass[fleqn,usenatbib]{mnras}
\usepackage[T1]{fontenc}
\usepackage{ae,aecompl}
\usepackage{multicol}
\usepackage{longtable}
\usepackage{supertabular}
\usepackage{epsfig}
\usepackage{lineno}
\usepackage{lscape}
\usepackage{rotating}
\usepackage{natbib}
\usepackage{graphics}
\usepackage{graphicx}    
\usepackage{amsmath}    
\usepackage[nolist,nohyperlinks]{acronym}
\usepackage{amssymb}    
\usepackage{hyperref}
\usepackage{float}
\hypersetup{draft}
\newcommand{\HI}{H\,{\sc i}} 

\title[The distance to MAXI J1348--630]{Measuring the distance to the black hole candidate X-ray binary MAXI J1348--630 using H\,{\sc i} absorption}

\author[J. Chauhan et al.]{J. Chauhan$^{1}$\thanks{E-mail: j.chauhan@student.curtin.edu.au (JC)}, J.~C.~A. Miller-Jones$^{1}$\thanks{james.miller-jones@curtin.edu.au (JCAM-J)}, W. Raja$^{2}$, J.~R. Allison$^{3}$, P.~F.~L. Jacob$^{3}$, 
\newauthor G.~E. Anderson$^{1}$, F. Carotenuto$^{4}$, S. Corbel$^{4,5}$, R. Fender$^{3}$, A. Hotan$^{2}$, M. Whiting$^{2}$,
\newauthor P.~A. Woudt$^{6}$, B. Koribalski$^{2}$, E. Mahony$^{2}$\\
$^{1}$International Centre for Radio Astronomy Research -- Curtin University, GPO Box U1987, Perth, WA 6845, Australia\\
$^{2}$CSIRO Astronomy and Space Science, Australia Telescope National Facility, PO Box 76, Epping NSW 1710, Australia\\
$^{3}$Sub-Dept. of Astrophysics, Department of Physics, University of Oxford, Denys Wilkinson Building, Keble Rd., Oxford, OX1 3RH, UK\\
$^{4}$AIM, CEA, CNRS, Universite{\'\i} Paris Diderot, Sorbonne Paris Cite{\'\i}, Universite{\'\i} Paris-Saclay, F-91191 Gif-sur-Yvette, France\\
$^{5}$Station de Radioastronomie de Nançay, Observatoire de Paris, PSL Research University, CNRS, Univ. Orl{\'e}ans, 18330 Nançay, France\\
$^{6}$Department of Astronomy, University of Cape Town, Private Bag X3, Rondebosch 7701, South Africa}

\date{Accepted XXX. Received YYY; in original form ZZZ}

\pubyear{2020}

\begin{document}
\label{firstpage}
\pagerange{\pageref{firstpage}--\pageref{lastpage}}
\maketitle

\begin{acronym}
    \acro{XRB}{X-ray binary}
    \acrodefplural{XRB}{X-ray binaries}
\end{acronym}

\begin{abstract}
We present H\,{\sc i} absorption spectra of the black hole candidate X-ray binary (XRB) MAXI J1348--630 using the Australian Square Kilometre Array Pathfinder (ASKAP) and MeerKAT. The ASKAP H\,{\sc i} spectrum shows a maximum negative radial velocity (with respect to the local standard of rest) of $-31\pm4$ km s$^{-1}$ for MAXI J1348--630, as compared to $-50\pm4$\,km\,s$^{-1}$ for a stacked spectrum of several nearby extragalactic sources. This implies a most probable distance of $2.2^{+0.5}_{-0.6}$\,kpc for MAXI J1348--630, and a strong upper limit of the tangent point distance at $5.3\pm0.1$\,kpc. Our preferred distance implies that MAXI J1348--630 reached $17\pm10$\,\% of the Eddington luminosity at the peak of its outburst, and that the source transited from the soft to the hard X-ray spectral state at $2.5\pm1.5$\,\% of the Eddington luminosity. The MeerKAT H\,{\sc i} spectrum of MAXI J1348--630 (obtained from the older, low-resolution 4k mode) is consistent with the re-binned ASKAP spectrum, highlighting the potential of the eventual capabilities of MeerKAT for XRB spectral line studies.
\end{abstract}

\begin{keywords}
black hole physics -- ISM: jets and outflows -- radio continuum: transients -- X-rays: binaries -- X-rays: individual: MAXI J1348--630
\vspace{-3.0em}
\end{keywords}

\acresetall

\renewcommand{\thefootnote}{\arabic{footnote}}
\section{Introduction}
\vspace{-0.5em}
The distance to an astrophysical object is a key physical parameter, as it provides us with a way to determine physical quantities from observables. For X-ray binaries (XRBs) in our Galaxy, the distance can be directly determined by measuring the parallax to the source, with either very long baseline interferometry \citep[VLBI; e.g.][]{Miller-Jones2009, Reid2011, Atri2020} or {\it Gaia} \citep{Gaia2018}. XRB distances can also be determined by studying the line features in infrared/optical spectra of the companion star \citep{Jonker2004}, by using the proper motions of the jet ejecta \citep{Mirabel1994} or by studying the interstellar extinction \citep{Zdziarski1998}.  However, these distances are often poorly constrained, and suffer large uncertainties that in some cases can exceed 50\% \citep{Jonker2004}. If the donor star is too faint, or a parallax measurement is not possible, an alternative technique is to use the 21-cm line of neutral hydrogen \citep[{\HI}, e.g.][]{Goss1977, Dickey1983, Dhawan2000, Chauhan2019b}. The Galactic H\,{\sc i} clouds along the line of sight to an XRB are rotating about the centre of the Milky Way, and their Doppler-shifted absorption features allow us to determine the kinematic distance to the source  \citep{Gathier1986, Kuchar1990}. While the kinematic distance estimates obtained from this method are fairly accurate (at least for circular rotation) when the source is located beyond the Solar orbit, they suffer ambiguities if the source is located within the Solar circle.  Since the rotation curves are then dual-valued, this leads to two distance estimates (near and far) for a single velocity measurement \citep[e.g.][]{Reid2014, Wenger2018}.

On 2019 January 26, the Monitor of All-sky X-ray Image\footnote{\url{http://maxi.riken.jp/top/index.html}} \citep[{\textit{MAXI}};][]{Matsuoka2009} discovered an uncatalogued XRB at a position RA (J2000) = 13$^{\rm h}$48$^{\rm m}$12\fs79$\pm0.03$ Dec (J2000) = $-63^{\rm d}$16$^{\prime}$28\farcs48$\pm0.04$ \citep[Galactic coordinates $l=309\fdg26407$, $b=-1\fdg10297$;][]{Russell2019a}, referred to as MAXI J1348--630 \citep{Yatabe2019}. The optical counterpart of the source was detected by \citet{Denisenko2019}, and the outburst was subsequently observed across the electromagnetic spectrum \citep[e.g.][]{Carotenuto2019, Chauhan2019a, Kennea2019}. The system is believed to harbour a stellar-mass black hole \citep{Russell2019a, LZhang2020}, although the key system parameters are poorly constrained.

In this investigation, we present H\,{\sc i} absorption spectra from both ASKAP and MeerKAT observations, and use the Doppler-shifted 21-cm absorption line to constrain the distance to MAXI J1348--630. We further use our distance constraints to estimate the peak X-ray luminosity and the spectral state transition luminosity.

\vspace{-2.0em}
\section{Observations and Data reduction}
\vspace{-0.5em}
\subsection{ASKAP} \label{sec:askap_Obs}
\vspace{-0.5em}
ASKAP \citep{Hotan2014} observed MAXI J1348--630 on 2019 February 13 for 9.91 hours (on source exposure 8.39 hours) from 13:22--23:16 UTC, using the full array of 36 dishes, with 36 overlapping beams. The large field of view ($\approx$ 30 deg$^{2}$) and high angular resolution ($\sim25^{\prime\prime}$) of ASKAP allow us to simultaneously detect the H\,{\sc i} absorption towards both MAXI J1348--630 and a set of nearby extragalactic sources (Table~\ref{tab:tab1}), enabling a discrimination between near and far kinematic distances. Our observation was performed at a central frequency of 1.34\,GHz, with a total bandwidth of 288\,MHz divided into 15368 fine channels, each of which has a frequency resolution of 18.519 kHz (velocity resolution 3.9\,km\,s$^{-1}$).

\begin{table}
\vspace{-1.8em}
 \centering
 \caption{Co-ordinates and 1.42-GHz ASKAP flux densities of MAXI J1348--630 and our extragalactic comparison sources. Positions taken from $[1]$ \citet{Russell2019a}; $[2]$ \citet{Murphy2007}.} 
 \vspace{-2.0em}
\begin{center}
\scalebox{0.8}{%
\begin{tabular}{ |l|c|c|r| }
 \hline
\hline
Source & RA (J2000) & Dec (J2000) & Flux Density$^{\mathrm{a}}$\\
Name & (hh:mm:ss) & (dd:mm:ss) & (mJy)\\
\hline
MAXI J1348--630 ${[1]}$ & 13:48:12.79 & --63:16:28.48 & $\phantom{1}155\pm2$\\
MGPS J134353--624941 ${[2]}$ & 13:43:53.13 & --62:49:41.4 & $\phantom{2}86\pm1$\\
MGPS J134551--634755 ${[2]}$ & 13:45:51.50 & --63:47:55.7 & $\phantom{1}104\pm1$\\
MGPS J134559--635023 ${[2]}$ & 13:45:59.85 & --63:50:23.1 & $\phantom{2}79\pm1$\\
MGPS J134625--632600 ${[2]}$ & 13:46:25.73 & --63:26:00.7 & $\phantom{2}71\pm1$\\
MGPS J135145--635836 ${[2]}$ & 13:51:45.91 & --63:58:36.8 & $\phantom{1}114\pm1$\\
MGPS J135236--631600 ${[2]}$ & 13:52:36.51 & --63:16:00.0 & $\phantom{1}141\pm1$\\
MGPS J135401--633032 ${[2]}$ & 13:54:01.09 & --63:30:32.1 & $\phantom{2}81\pm1$\\
MGPS J135546--632642 ${[2]}$ & 13:55:46.25 & --63:26:42.5 & $1181\pm2$\\
\hline
\hline
\end{tabular}}\\
\end{center}
\begin{flushleft}
$^{\mathrm{a}}$ $1\sigma$ errors are quoted, calculated by adding in quadrature the error on the Gaussian fit and the rms noise in the image.\\
\end{flushleft}
\label{tab:tab1}
\vspace{-1.0em}
\end{table}
For reducing our multiple-beam full array data on MAXI J1348--630, we used the ASKAP data analysis software, {\tt ASKAPsoft}\footnote{\url{http://www.atnf.csiro.au/computing/software/askapsoft/sdp/docs/current/index.html}}. Although our MAXI J1348--630 data have more antennas and beams, we follow a similar calibration procedure to that described in \citet{Chauhan2019b}. However, for generating the spectral cube from the calibrated, continuum-subtracted measurement set created by {\tt ASKAPsoft}, we used the {\tt TCLEAN} task in the Common Astronomy Software Application \citep[{\tt CASA} v5.1.2-4:][]{McMullin2007} to ensure we could use the same procedure to extract both MeerKAT and ASKAP spectra, and to allow quick optimisation of our imaging parameters (e.g.\ {\it uv}-range, deconvolution depth, deconvolver), given the demand on supercomputing time to run {\tt ASKAPsoft}. We produced a spectral sub-cube of 378 channels centered at the rest frequency (1420.4 MHz) of the H\,{\sc i} line using a Briggs weighting parameter (robustness) of 0.5, and adopting a minimum baseline length of 700\,m.

From the ASKAP spectral cube, we extracted the H\,{\sc i} absorption spectrum for MAXI J1348--630 and the eight extragalactic sources listed in Table \ref{tab:tab1}, by measuring the brightness in each frequency channel at the position corresponding to the peak flux density (determined from the continuum image). We used the {\tt IMFIT} task in {\tt CASA} to measure source flux densities and $1\sigma$ uncertainties from the continuum images.

\vspace{-1.5em}
\subsection{MeerKAT}
\vspace{-0.5em}
MAXI J1348--630 was monitored as part of the MeerKAT Large Survey Project for slow transients (ThunderKAT; \citealt{Fender2016}). Here we use data from 2019 February 09 between 05:08 and 05:23 (UTC), when the source was brightest in the radio ($486\pm2$\,mJy at 1.42 GHz; Carotenuto et al., in prep.) and therefore most sensitive to \mbox{H\,{\sc i}} absorption. MeerKAT provides a field of view of 0.86 deg$^{2}$, and a spatial resolution of $\sim10^{\prime\prime}$ at 1.42 GHz. Our observations were carried out using 60 MeerKAT antennas, at a central frequency of 1284\,MHz, with 860\,MHz of bandwidth. Only the 4k correlator mode was available, which gave a spectral resolution of 209\,kHz, equal to 44\,km\,s$^{-1}$ at 1420.4\,MHz.

Spectral line data reduction was carried out using a standard procedure that implemented tasks from {\tt MIRIAD} (\citealt{Sault1995}). The data were first converted to FITS format using {\tt CASA}, selecting only channels in the range 10\,MHz either side of 1420.4\,MHz (equivalent to radial velocities of $\pm 2000$\,km\,s$^{-1}$). Calibration of the bandpass and flux scale was carried out using PKS B1934--638 (\citealt{Reynolds:1994}), and the time varying antenna gains using PKS B1421--490. To avoid corrupting the calibration solutions with Galactic \mbox{H\,{\sc i}} emission and absorption, the central 20 channels ($\pm 400$\,km\,s$^{-1}$) were flagged and then interpolated from neighbouring channels. Further self-calibration of the MAXI J1348--630 data was carried out to correct the time-varying gain phase. After subtraction of the continuum flux density, a spectral cube was formed within 5\,arcmin of MAXI J1348--630 using a robustness of 0.5, and a minimum baseline length of 700\,m. The final spectrum was extracted from the cube, adopting a similar procedure to that described for ASKAP in Section \ref{sec:askap_Obs}.

\vspace{-2.0em}
\section{Results}
\begin{figure*}
\vspace{-2.5em}
    \includegraphics[width=\columnwidth]{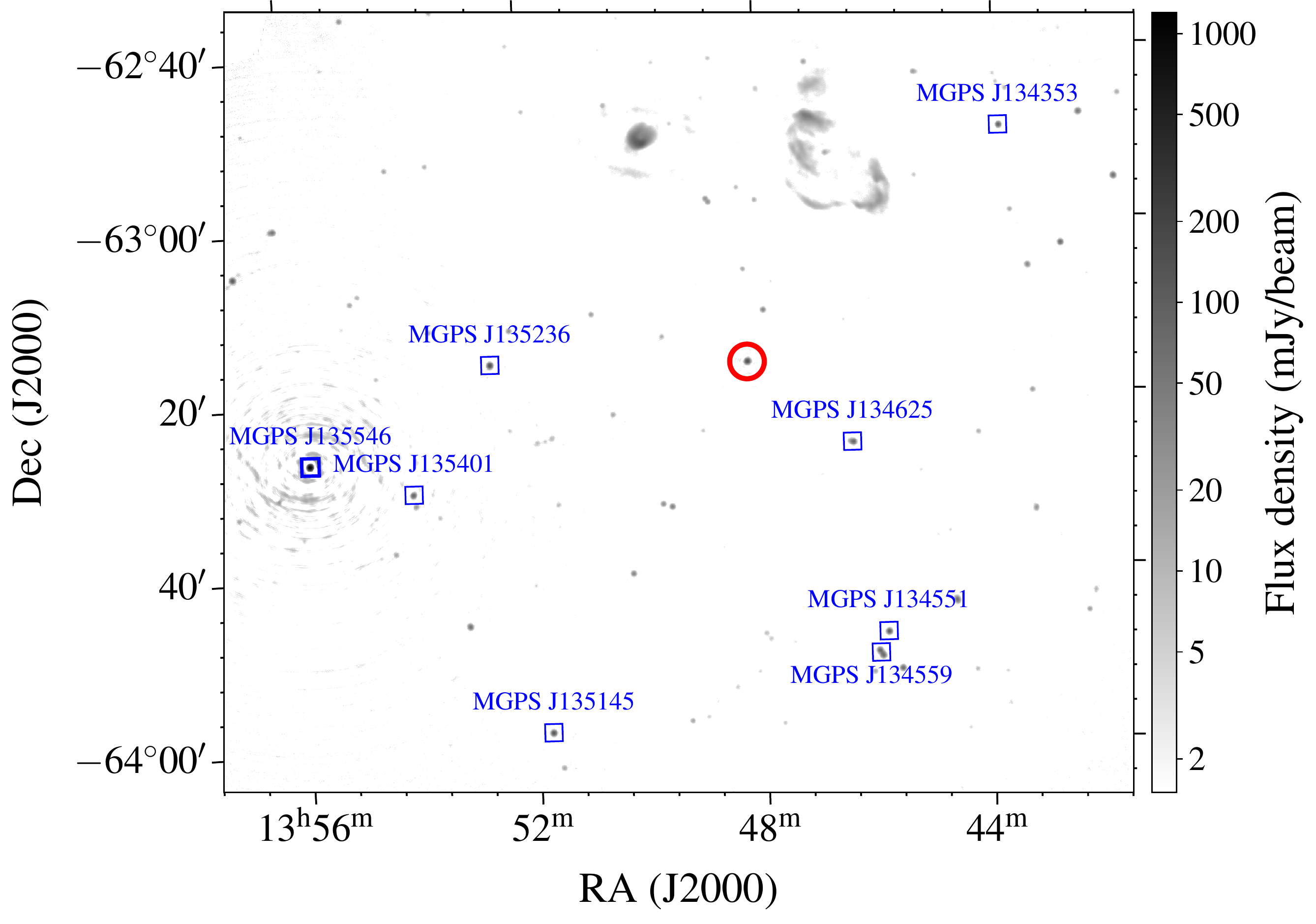}
    \includegraphics[width=\columnwidth]{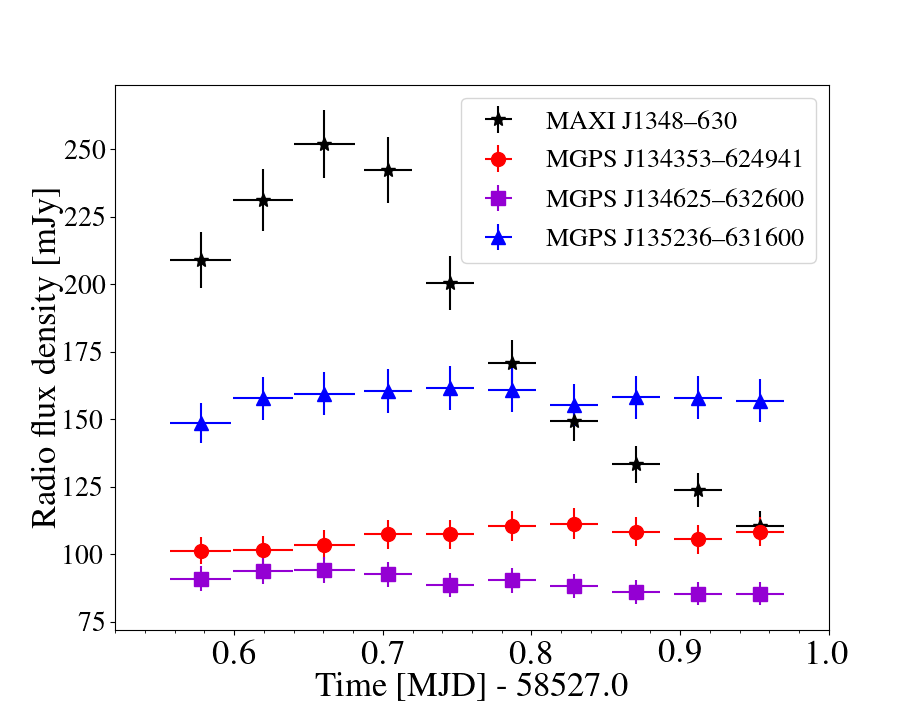}\vspace{-1.0em}
    \caption{{\textit{Left}} panel: A continuum mosaic of the field surrounding MAXI J1348--630, from our 1.34-GHz ASKAP observation on 2019 February 13. The image has a size of 1.7$^\circ$ $\times$ 1.5$^\circ$, centered at RA = 13$^{\rm h}$49$^{\rm m}$19\fs78, DEC = $-63^{\rm d}$20$^{\prime}$23\farcs93. The location of MAXI J1348--630 is highlighted with the red circle, and the comparison extragalactic sources (listed in Table \ref{tab:tab1}) are indicated by the blue squares. {\textit{Right}} panel: The time-resolved 1.34-GHz ASKAP light curve of MAXI J1348--630 (black stars) and the extragalactic sources, for beam 20 only. The flux density of MAXI J1348--630 varies on a $\sim1$--hour time scale, whereas the extragalactic sources remain constant within error bars, demonstrating that the variability observed in MAXI J1348--630 is intrinsic to the source.} 
    \vspace{-1.7em}
    \label{fig:fig1}
\end{figure*}

\vspace{-0.5em}
In the left panel of Fig.~\ref{fig:fig1} we present an ASKAP continuum image of the MAXI J1348--630 field created by mosaicing beams 14, 15, 20 and 21. Residual uncertainties remain around the bright ($>1$ Jy) sources (e.g.\ extragalactic source MGPS J135546--632642 in Fig.~\ref{fig:fig1}) at the level of 1--2\%, due to remaining calibration and deconvolution errors.

We detected MAXI J1348--630 in our ASKAP data with a flux density of $155\pm2$\,mJy at 1.42\,GHz. We also selected eight extragalactic comparison sources in the field, which are listed in Table \ref{tab:tab1}, and also shown in Fig.~\ref{fig:fig1} (left panel). At the time of the ASKAP observation, MAXI J1348--630 was transiting from the hard to the soft X-ray spectral state \citep{Tominaga2020}, where many black hole XRBs undergo transient jet ejection events \citep{Fender2004}.

To characterize the short-timescale variability in MAXI J1348--630 during our ASKAP observation, we generated a continuum image for each of ten hour-long time bins. We present the time-resolved 1.34-GHz light curve of MAXI J1348--630 for beam 20 (for which the source is closest to the centre of the beam) in the right panel of Fig.~\ref{fig:fig1}. This shows a short-duration flare, peaking at $252\pm13$ mJy at 15:51:32 (UTC), and then gradually decreasing to $111\pm6$\,mJy by 22:53:16 (UTC). The radio flux density of the extragalactic sources in the same ASKAP beam remained constant during the observation, verifying that the variation seen from MAXI J1348--630 is intrinsic to the source.

\vspace{-1.5em}
\subsection{H\,{\sc i} absorption spectra \label{sec:HIabsspec}}
\begin{figure}
   \includegraphics[width=\columnwidth]{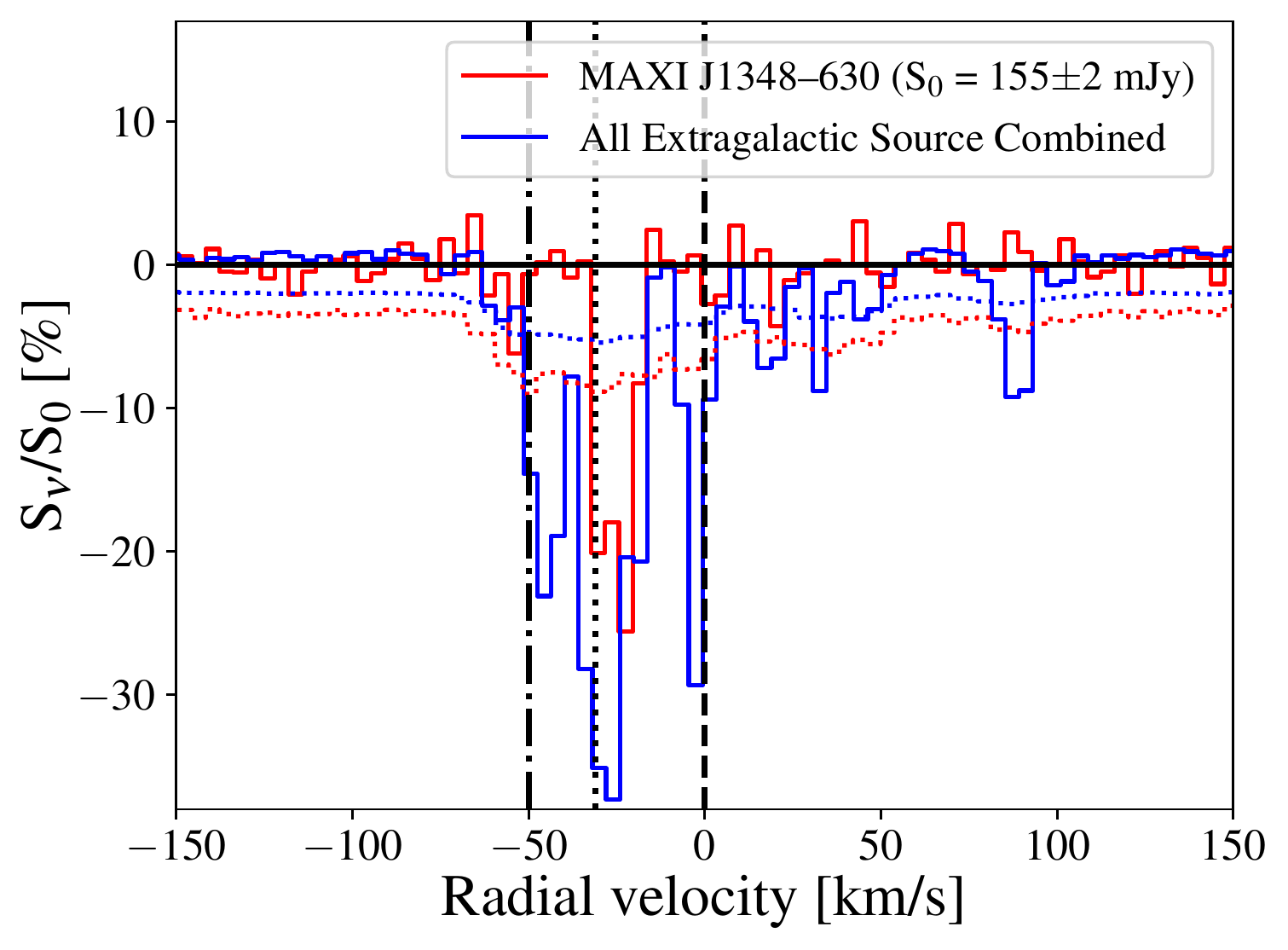}\vspace{-2.0em}
    \caption{The H\,{\sc i} absorption complex observed in the direction of MAXI J1348--630 using our ASKAP observation from 2019 February 13. $S_{\nu}$ is calculated from the spectral cube, whereas $S_{0}$ is measured from the continuum image. The red and blue curves show the H\,{\sc i} absorption against MAXI J1348--630 and the stack of the eight extragalactic sources (Table \ref{tab:tab1}), respectively. The corresponding dotted lines represent the respective per-channel $3\sigma$ noise levels (normalized to the source continuum flux density), taken from nearby source-free regions. The black dashed vertical line represents the rest frequency of the H\,{\sc i} line. The dotted ($-31$\,km\,s$^{-1}$) and dot-dashed vertical ($-50$\,km\,s$^{-1}$) lines highlight the ($>3\sigma$ significant) maximum negative radial velocities (with respect to the LSR) observed from the spectra of MAXI J1348--630, and the merged extragalactic sources, respectively. The maximum negative velocity of $-31\pm4$\,km\,s$^{-1}$ towards MAXI J1348--630 determines the most probable distance as $2.2^{+0.5}_{-0.6}$\,kpc, whereas the non-detection of more negative velocities sets a stringent upper limit of the tangent point at $5.3\pm0.1$\, kpc.
    }
    \label{fig:fig2}
\vspace{-1.0em}
\end{figure}

\vspace{-0.5em}
In Fig.~\ref{fig:fig2}, we show the spectrum for MAXI J1348--630, together with a stacked spectrum for all eight extragalactic sources, and the $3\sigma$ noise levels measured from nearby regions. We detect significant ($>3\sigma$) H\,{\sc i} absorption complexes out to maximum negative velocities (with respect to the local standard of rest, or LSR) of $-31\pm4$ km s$^{-1}$ and $-50\pm4$\,km\,s$^{-1}$ for MAXI J1348--630 and the extragalactic sources, respectively. 

\begin{figure}
    \includegraphics[width=0.9\columnwidth]{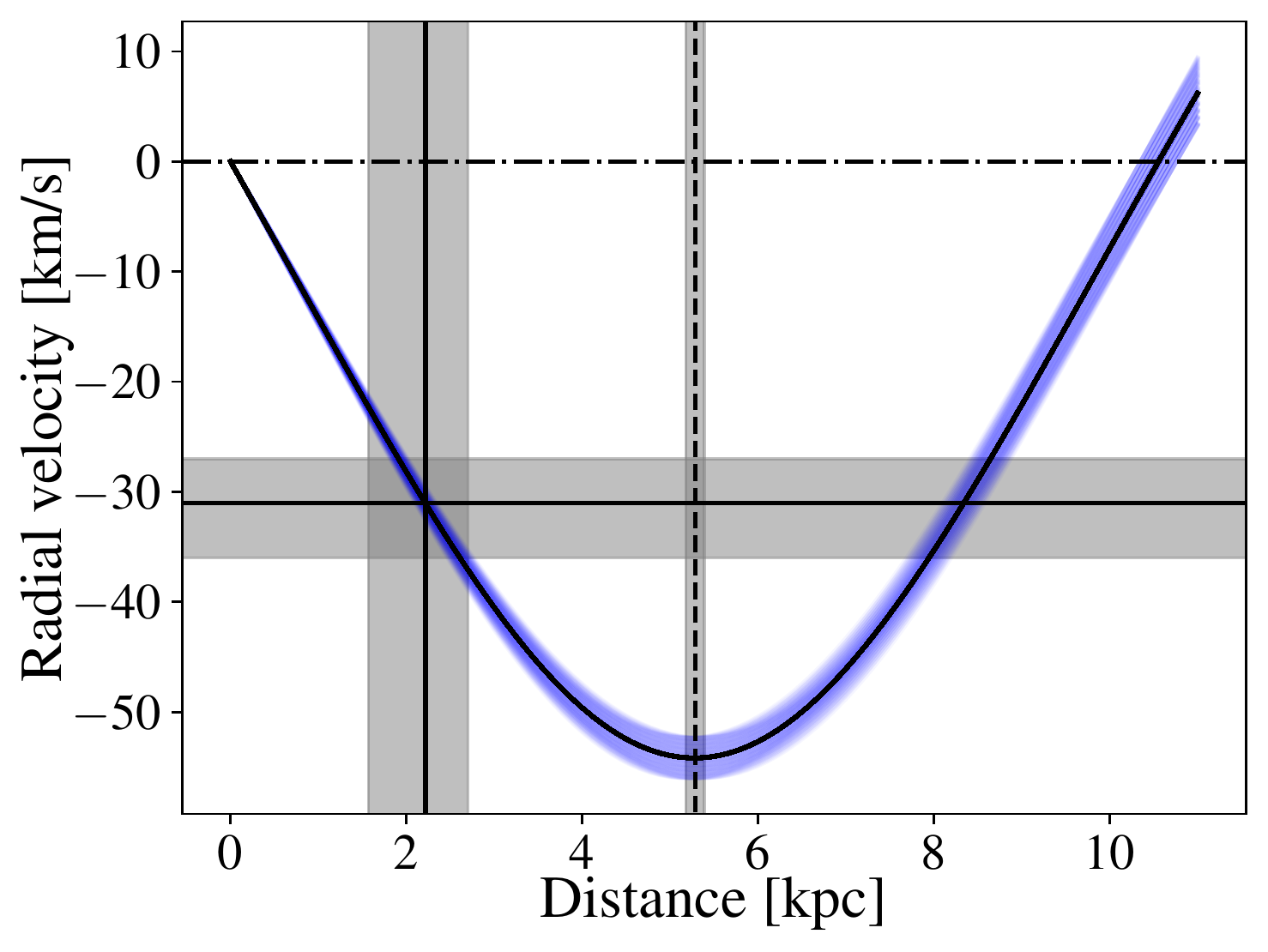}\vspace{-1.0em}
    \caption{The expected variation of the radial velocity of the local standard of rest in the direction of MAXI J1348--630 with distance from the Sun, calculated using the Monte Carlo approach described by \citet{Wenger2018}. The black line represents the expected curve, and the blue shaded region shows the effect of incorporating the $1\sigma$ uncertainties on the Galactic rotation parameters from \citet{Reid2014}. The horizontal solid line shows the maximum negative radial velocity (with respect to the LSR) measured from the H\,{\sc i} absorption spectrum. The solid and dashed vertical lines show the most likely distance and the tangent point distance, respectively. The grey shaded regions show the $1\sigma$ uncertainties. The most probable distance of MAXI J1348--630 is the near kinematic distance of $2.2^{+0.5}_{-0.6}$\,kpc.
    }
    \label{fig:fig4}
\vspace{-1.0em}
\end{figure}

To determine the distance to MAXI J1348--630, we computed the Galactic rotation curve for the Galactic longitude of MAXI J1348--630, and determined the variation of the radial velocity ($V_{\rm LSR}$) of the LSR with distance from the Sun ($d$).  For simplicity we assumed that the Galactic rotation curve is flat, that the Milky Way is rotating with a circular velocity ($V_{0}$) of $240\pm8$\,km\,s$^{-1}$ \citep{Reid2014}, and that the Sun is located at a distance $R_{0} = 8.34\pm0.16$\,kpc from the Galactic centre \citep{Reid2014}. Fig.~\ref{fig:fig4} shows that the predicted LSR radial velocities are negative within a few kpc of the Sun, and by comparison with our observed H\,{\sc i} spectrum allows us to determine the near distance, the far distance and the tangent point distance ($R_{T} = R_{0}\cos{l}\equiv5.3\pm0.1$\,kpc) for this line of sight.

To determine robust constraints on the kinematic distance, we adopted the recipes suggested by \citet{Wenger2018}, who developed a Monte Carlo approach\footnote{\url{http://www.treywenger.com/kd/index.php}} for inferring kinematic distances and associated uncertainties, using the rotation curve provided by \citet{Reid2014}. Using this approach, we determined the near and far distances of MAXI J1348--630 to be $2.2^{+0.5}_{-0.6}$\,kpc and $8.4^{+0.6}_{-0.6}$\,kpc, respectively. The maximum negative absorption velocity (with respect to the LSR) observed for the extragalactic sources ($-50\pm4$\,km\,s$^{-1}$) is in good agreement with the tangent point velocity ($-54\pm4$\,km\,s$^{-1}$) for this line of sight (within error bars, and calculated using \citealt{Reid2014} and \citealt{Wenger2018}). H\,{\sc i} absorption at the tangent point velocity is not observed towards MAXI J1348--630, implying that it must be closer than the tangent point distance of $5.3\pm0.1$\,kpc, and thereby most likely placing the source at the near kinematic distance of $2.2^{+0.5}_{-0.6}$\,kpc.

\vspace{-1.5em}
\subsubsection{Comparison with the MeerKAT spectrum}
\begin{figure}
\vspace{-2.0em}
    \includegraphics[width=1.0\columnwidth]{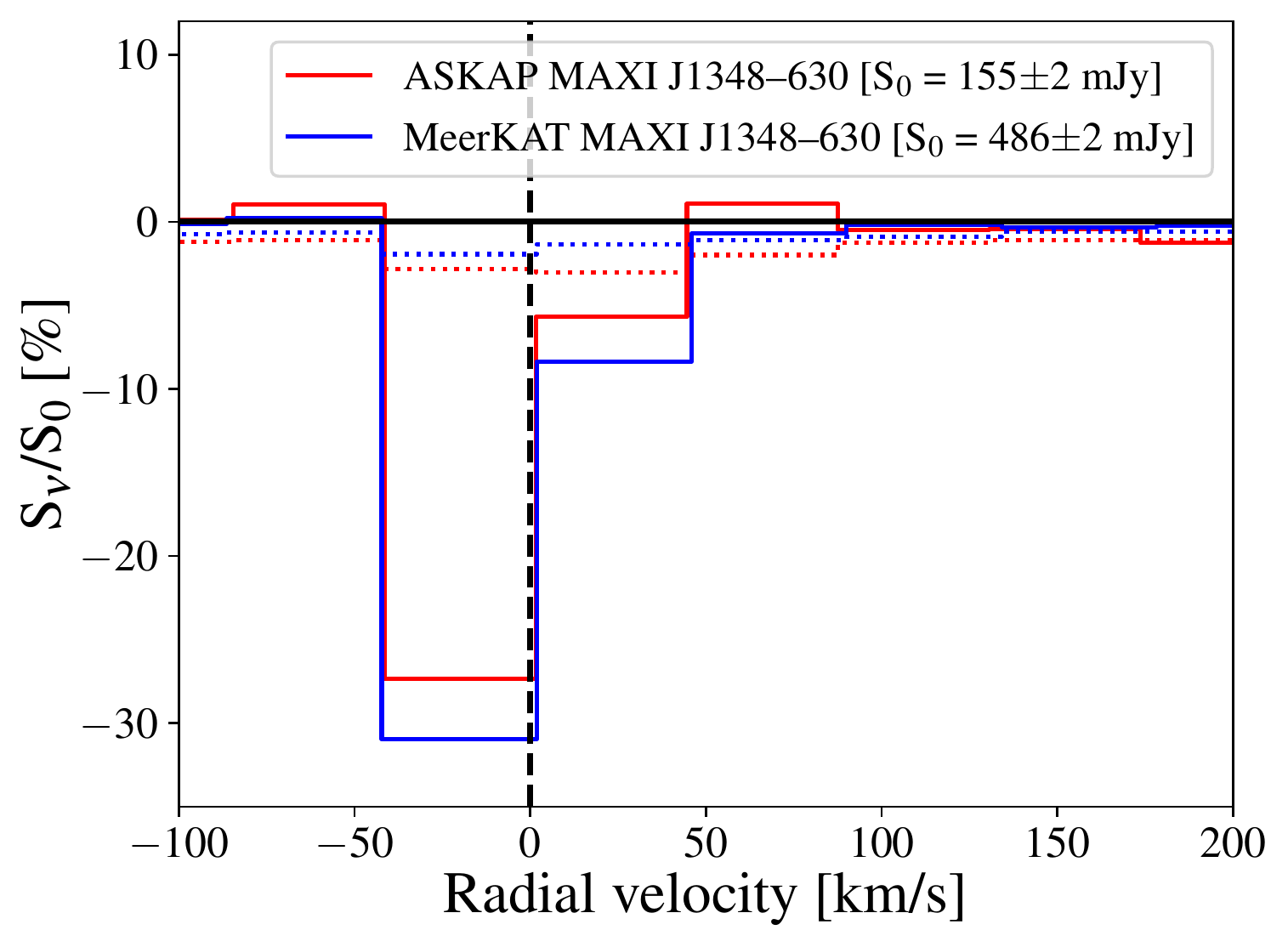}\vspace{-2.0em}
    \caption{The MeerKAT (blue) and rebinned ASKAP (red) H\,{\sc i} absorption spectra towards MAXI J1348--630. The $3\sigma$ rms noise levels for both instruments are shown as dotted lines. The two spectra match well within uncertainties.
    }
\vspace{-1.0em}
    \label{fig:fig3}
\end{figure}

\vspace{-0.5em}
MeerKAT detected MAXI J1348--630 as a bright point source, of flux density $486\pm2$\,mJy at 1.42 GHz (Carotenuto et al., in prep.). In the 4k correlator mode that was used, MeerKAT had a velocity resolution of 44\,km\,s$^{-1}$, so we rebinned our ASKAP data (with velocity resolution 3.9\,km\,s$^{-1}$) to match the MeerKAT resolution. The uncertainties on the MeerKAT spectra are smaller, both because the instrument has a lower system temperature, and because MAXI J1348--630 was brighter at the time of the MeerKAT observations 
than it was during the ASKAP observations ($155\pm2$\,mJy at 1.42 GHz). Nonetheless, as shown in Figure \ref{fig:fig3}, the two spectra are consistent within uncertainties.

\vspace{-2.0em}
\section{Discussion}
\vspace{-0.5em}
\subsection{Distance constraint and its implications}
\vspace{-0.5em}
\citet{Tominaga2020} studied the complete outburst of MAXI J1348--630 using {\it MAXI}/GSC data in the 2--20 keV energy range. They found the peak of the outburst to have occurred on 2019 February 09, and the soft-to-hard spectral state transition to have occurred on 2019 April 27 \citep{Tominaga2020}.  We used the High Energy Astrophysics Science Archive Research Center (HEASARC) tool WebPIMMS\footnote{\url{https://heasarc.gsfc.nasa.gov/cgi-bin/Tools/w3pimms/w3pimms.pl}} to calculate the bolometric X-ray flux in the energy range 0.01--100 keV from the X-ray flux (2--20 keV) and spectral model determined by \citet{Tominaga2020}. We derived a peak unabsorbed X-ray flux of $2.8\pm0.2\times10^{-7}$ erg\,cm$^{-2}$\,s$^{-1}$, and an unabsorbed flux of $4.2\pm0.5\times10^{-8}$ erg\,cm$^{-2}$\,s$^{-1}$ for the soft-to-hard X-ray spectral state transition, corresponding to luminosities of $1.6\pm0.9\times10^{38}$ and $2.4\pm1.4\times10^{37}$ erg\,s$^{-1}$, respectively, at our preferred distance.

If we consider the compact object to be a black hole \citep{Russell2019a, LZhang2020} of typical mass $8\pm1\,M_{\odot}$ \citep{Kreidberg2012}, the peak unabsorbed luminosity corresponds to $0.17\pm0.10 L_{\rm Edd}$, where $L_{\rm Edd}$ is the Eddington luminosity. This is in reasonable agreement with the range of $0.2$--$0.4L_{\rm Edd}$ found for canonical black hole XRBs \citep{McClintock2009}. We further found that the system transitioned from the soft to the hard X-ray spectral state at $0.025\pm0.015L_{\rm Edd}$, consistent with the range of $0.003-0.03L_{\rm Edd}$ determined by \citet{Maccarone2003}, \citet{Kalemci2013} and \citet{Vahdat2019} for typical black hole XRBs.

Using the soft-to-hard X-ray spectral state transition luminosity and the measured column density towards the source, \citet{Tominaga2020} placed it in front of the Scutum-Centaurus arm, at $<4$\,kpc.  Our measured H\,{\sc i} distance is consistent with (albeit more precise than) this estimate, and implies a state transition luminosity towards the lower end of their assumed range.  \citet{Tominaga2020} further used their measured inner disk radius in the soft X-ray spectral state to constrain the black hole mass as a function of distance, inclination angle and black hole spin.  Fitting the X-ray data of \citet{Tominaga2020} with the same {\tt kerrbb} model, and leaving the distance free to vary within our $1\sigma$ uncertainty range, we find that the lower limit on the inferred black hole mass (for the case of a non-rotating black hole with an inclination angle of $0^{\circ}$) reduces to $3.7M_{\odot}$, rising to $5.8M_{\odot}$ for an inclination angle of $60^{\circ}$. If the system is at low inclination and slowly rotating, this weakens the evidence for a particularly massive black hole in the system.

\vspace{-1.5em}
\subsection{Future X-ray binary monitoring}
\vspace{-0.5em}
Over the past year, MeerKAT has conducted weekly XRB monitoring \citep[e.g.][]{Bright2020, Tremou2020, Williams2020} under the large survey project ThunderKAT \citep{Fender2016}. However, the low spectral resolution limited the potential for H\,{\sc i} studies. The recent correlator upgrade (providing 32K spectral channels with a velocity resolution of 6.1 km\,s$^{-1}$) enables spectral resolution comparable to that of ASKAP.  Our data demonstrate the future potential of MeerKAT, which with its high sensitivity will be well placed to routinely provide {\HI} distance estimates for all bright, outbursting XRBs.

The combined temporal coverage of our ASKAP and MeerKAT observations shows that the flare detected by ASKAP
(Fig.~\ref{fig:fig1}) is a secondary re-brightening, following the peak detected during the MeerKAT observation \citep{Carotenuto2019}. This is not unusual, as multiple jet ejections have been observed in several previous black hole XRB outbursts \citep[e.g.][]{Mirabel1994, Brocksopp2013}. In future, the combination of MeerKAT and ASKAP observations, together with higher frequency facilities, will provide high-cadence light curves that can help constrain key parameters such as jet speed, energetics and geometry \citep{Tetarenko2017}.

\vspace{-2.0em}
\section{Conclusions}
\label{sec:conclusions}
\vspace{-0.5em}
We have used ASKAP to detect H\,{\sc i} absorption towards MAXI J1348--630 out to a maximum negative radial velocity (with respect to the LSR) of $-31\pm4$ km s$^{-1}$, implying a most probable kinematic distance of $2.2^{+0.5}_{-0.6}$\,kpc. By comparison with the absorption towards a stack of the extragalactic sources in the field of view, we place a robust upper limit of $5.3\pm0.1$\, kpc, corresponding to the tangent point distance. 

Using our preferred distance and assuming a canonical black hole mass of $8\pm1M_{\sun}$, we found that MAXI J1348--630 was accreting at $17\pm10$\,\% of the Eddington luminosity over the peak of the outburst, and the soft-to-hard X-ray spectral state transition happened at  $2.5\pm1.5$\,\% of the Eddington luminosity, consistent in both cases with what has been found for other black hole XRBs.

Finally, our study highlights the synergies between ASKAP and MeerKAT, demonstrating the potential for routine H\,{\sc i} distance measurements for future black hole X-ray binaries in outburst.

\vspace{-2.0em}
\section*{Acknowledgements}
\vspace{-0.5em}
We thank the referee for their valuable comments. The Australian SKA Pathfinder is part of the Australia Telescope National Facility which is managed by CSIRO. Operation of ASKAP is funded by the Australian Government with support from the National Collaborative Research Infrastructure Strategy. ASKAP uses the resources of the Pawsey Supercomputing Centre. Establishment of ASKAP, the Murchison Radio-astronomy Observatory and the Pawsey Supercomputing Centre are initiatives of the Australian Government, with support from the Government of Western Australia and the Science and Industry Endowment Fund. We acknowledge the Wajarri Yamatji people as the traditional owners of the Observatory site. The MeerKAT telescope is operated by the South African Radio Astronomy Observatory, which is a facility of the National Research Foundation, an agency of the Department of Science and Innovation. JCAM-J and GEA are the recipients of an Australian Research Council Future Fellowship (FT140101082) and a Discovery Early Career Researcher Award (DE180100346), respectively, funded by the Australian Government.
\vspace{-2.0em}
\section*{Data Availability}
\vspace{-0.5em}
The processed ASKAP data are available in the CASDA archive at \url{ https://data.csiro.au/collections/}. The uncalibrated MeerKAT visibility data are publicly available, and stored at the South African Radio Astronomy Observatory (SARAO) Archive at \url{https://archive@ska.ac.za}.

\vspace{-2.0em}
\bibliographystyle{mnras}
\bibliography{References}

\begin{thebibliography}{}
\makeatletter
\relax
\def\mn@urlcharsother{\let\do\@makeother \do\$\do\&\do\#\do\^\do\_\do\%\do\~}
\def\mn@doi{\begingroup\mn@urlcharsother \@ifnextchar [ {\mn@doi@}
  {\mn@doi@[]}}
\def\mn@doi@[#1]#2{\def\@tempa{#1}\ifx\@tempa\@empty \href
  {http://dx.doi.org/#2} {doi:#2}\else \href {http://dx.doi.org/#2} {#1}\fi
  \endgroup}
\def\mn@eprint#1#2{\mn@eprint@#1:#2::\@nil}
\def\mn@eprint@arXiv#1{\href {http://arxiv.org/abs/#1} {{\tt arXiv:#1}}}
\def\mn@eprint@dblp#1{\href {http://dblp.uni-trier.de/rec/bibtex/#1.xml}
  {dblp:#1}}
\def\mn@eprint@#1:#2:#3:#4\@nil{\def\@tempa {#1}\def\@tempb {#2}\def\@tempc
  {#3}\ifx \@tempc \@empty \let \@tempc \@tempb \let \@tempb \@tempa \fi \ifx
  \@tempb \@empty \def\@tempb {arXiv}\fi \@ifundefined
  {mn@eprint@\@tempb}{\@tempb:\@tempc}{\expandafter \expandafter \csname
  mn@eprint@\@tempb\endcsname \expandafter{\@tempc}}}

\bibitem[\protect\citeauthoryear{{Atri} et~al.,}{{Atri}
  et~al.}{2020}]{Atri2020}
{Atri} P.,  et~al., 2020, \mn@doi [\mnras] {10.1093/mnrasl/slaa010}, \href
  {https://ui.adsabs.harvard.edu/abs/2020MNRAS.493L..81A} {493, L81}

\bibitem[\protect\citeauthoryear{{Bright} et~al.,}{{Bright}
  et~al.}{2020}]{Bright2020}
{Bright} J.~S.,  et~al., 2020, \mn@doi [Nature Astronomy]
  {10.1038/s41550-020-1023-5}, \href
  {https://ui.adsabs.harvard.edu/abs/2020NatAs...4..697B} {4, 697}

\bibitem[\protect\citeauthoryear{{Brocksopp}, {Corbel}, {Tzioumis},
  {Broderick}, {Rodriguez}, {Yang}, {Fender}  \& {Paragi}}{{Brocksopp}
  et~al.}{2013}]{Brocksopp2013}
{Brocksopp} C.,  {Corbel} S.,  {Tzioumis} A.,  {Broderick} J.~W.,  {Rodriguez}
  J.,  {Yang} J.,  {Fender} R.~P.,   {Paragi} Z.,  2013, \mn@doi [\mnras]
  {10.1093/mnras/stt493}, \href
  {https://ui.adsabs.harvard.edu/abs/2013MNRAS.432..931B} {432, 931}

\bibitem[\protect\citeauthoryear{{Carotenuto}, {Tremou}, {Corbel}, {Fender},
  {Woudt}  \& {Miller-Jones}}{{Carotenuto} et~al.}{2019}]{Carotenuto2019}
{Carotenuto} F.,  {Tremou} E.,  {Corbel} S.,  {Fender} R.,  {Woudt} P.,
  {Miller-Jones} J.,  2019, ATel, \href
  {https://ui.adsabs.harvard.edu/abs/2019ATel12497....1C} {12497, 1}

\bibitem[\protect\citeauthoryear{{Chauhan} et~al.,}{{Chauhan}
  et~al.}{2019a}]{Chauhan2019b}
{Chauhan} J.,  et~al., 2019a, \mn@doi [\mnras] {10.1093/mnrasl/slz113}, \href
  {https://ui.adsabs.harvard.edu/abs/2019MNRAS.488L.129C} {488, L129}

\bibitem[\protect\citeauthoryear{{Chauhan}, {Miller-Jones}, {Anderson},
  {Russell}, {Hancock}, {Bahramian}, {Duchesne}  \& {Williams}}{{Chauhan}
  et~al.}{2019b}]{Chauhan2019a}
{Chauhan} J.,  {Miller-Jones} J.,  {Anderson} G.,  {Russell} T.,  {Hancock} P.,
   {Bahramian} A.,  {Duchesne} S.,   {Williams} A.,  2019b, ATel, \href
  {https://ui.adsabs.harvard.edu/abs/2019ATel12520....1C} {12520, 1}

\bibitem[\protect\citeauthoryear{{Denisenko} et~al.,}{{Denisenko}
  et~al.}{2019}]{Denisenko2019}
{Denisenko} D.,  et~al., 2019, ATel, \href
  {https://ui.adsabs.harvard.edu/abs/2019ATel12430....1D} {12430, 1}

\bibitem[\protect\citeauthoryear{{Dhawan}, {Goss}  \&
  {Rodr{\'\i}guez}}{{Dhawan} et~al.}{2000}]{Dhawan2000}
{Dhawan} V.,  {Goss} W.~M.,   {Rodr{\'\i}guez} L.~F.,  2000, \mn@doi [\apj]
  {10.1086/309371}, \href
  {https://ui.adsabs.harvard.edu/abs/2000ApJ...540..863D} {540, 863}

\bibitem[\protect\citeauthoryear{{Dickey}}{{Dickey}}{1983}]{Dickey1983}
{Dickey} J.~M.,  1983, \mn@doi [\apjl] {10.1086/184132}, \href
  {https://ui.adsabs.harvard.edu/abs/1983ApJ...273L..71D} {273, L71}

\bibitem[\protect\citeauthoryear{{Fender}, {Belloni}  \& {Gallo}}{{Fender}
  et~al.}{2004}]{Fender2004}
{Fender} R.~P.,  {Belloni} T.~M.,   {Gallo} E.,  2004, \mn@doi [\mnras]
  {10.1111/j.1365-2966.2004.08384.x}, \href
  {https://ui.adsabs.harvard.edu/abs/2004MNRAS.355.1105F} {355, 1105}

\bibitem[\protect\citeauthoryear{{Fender} et~al.,}{{Fender}
  et~al.}{2016}]{Fender2016}
{Fender} R.,  et~al., 2016, in MeerKAT Science: On the Pathway to the SKA.
  p.~13 (\mn@eprint {arXiv} {1711.04132})

\bibitem[\protect\citeauthoryear{{Gaia Collaboration} et~al.,}{{Gaia
  Collaboration} et~al.}{2018}]{Gaia2018}
{Gaia Collaboration} et~al., 2018, \mn@doi [\aap]
  {10.1051/0004-6361/201833051}, \href
  {https://ui.adsabs.harvard.edu/abs/2018A&A...616A...1G} {616, A1}

\bibitem[\protect\citeauthoryear{{Gathier}, {Pottasch}  \& {Goss}}{{Gathier}
  et~al.}{1986}]{Gathier1986}
{Gathier} R.,  {Pottasch} S.~R.,   {Goss} W.~M.,  1986, \aap, \href
  {https://ui.adsabs.harvard.edu/abs/1986A&A...157..191G} {157, 191}

\bibitem[\protect\citeauthoryear{{Goss} \& {Mebold}}{{Goss} \&
  {Mebold}}{1977}]{Goss1977}
{Goss} W.~M.,  {Mebold} U.,  1977, \mn@doi [\mnras] {10.1093/mnras/181.2.255},
  \href {https://ui.adsabs.harvard.edu/abs/1977MNRAS.181..255G} {181, 255}

\bibitem[\protect\citeauthoryear{{Hotan} et~al.,}{{Hotan}
  et~al.}{2014}]{Hotan2014}
{Hotan} A.~W.,  et~al., 2014, \mn@doi [\pasa] {10.1017/pasa.2014.36}, \href
  {https://ui.adsabs.harvard.edu/abs/2014PASA...31...41H} {31, e041}

\bibitem[\protect\citeauthoryear{{Jonker} \& {Nelemans}}{{Jonker} \&
  {Nelemans}}{2004}]{Jonker2004}
{Jonker} P.~G.,  {Nelemans} G.,  2004, \mn@doi [\mnras]
  {10.1111/j.1365-2966.2004.08193.x}, \href
  {https://ui.adsabs.harvard.edu/abs/2004MNRAS.354..355J} {354, 355}

\bibitem[\protect\citeauthoryear{{Kalemci}, {Din{\c{c}}er}, {Tomsick},
  {Buxton}, {Bailyn}  \& {Chun}}{{Kalemci} et~al.}{2013}]{Kalemci2013}
{Kalemci} E.,  {Din{\c{c}}er} T.,  {Tomsick} J.~A.,  {Buxton} M.~M.,  {Bailyn}
  C.~D.,   {Chun} Y.~Y.,  2013, \mn@doi [\apj] {10.1088/0004-637X/779/2/95},
  \href {https://ui.adsabs.harvard.edu/abs/2013ApJ...779...95K} {779, 95}

\bibitem[\protect\citeauthoryear{{Kennea} \& {Negoro}}{{Kennea} \&
  {Negoro}}{2019}]{Kennea2019}
{Kennea} J.~A.,  {Negoro} H.,  2019, ATel, \href
  {https://ui.adsabs.harvard.edu/abs/2019ATel12434....1K} {12434, 1}

\bibitem[\protect\citeauthoryear{{Kreidberg}, {Bailyn}, {Farr}  \&
  {Kalogera}}{{Kreidberg} et~al.}{2012}]{Kreidberg2012}
{Kreidberg} L.,  {Bailyn} C.~D.,  {Farr} W.~M.,   {Kalogera} V.,  2012, \mn@doi
  [\apj] {10.1088/0004-637X/757/1/36}, \href
  {https://ui.adsabs.harvard.edu/abs/2012ApJ...757...36K} {757, 36}

\bibitem[\protect\citeauthoryear{{Kuchar} \& {Bania}}{{Kuchar} \&
  {Bania}}{1990}]{Kuchar1990}
{Kuchar} T.~A.,  {Bania} T.~M.,  1990, \mn@doi [\apj] {10.1086/168527}, \href
  {https://ui.adsabs.harvard.edu/abs/1990ApJ...352..192K} {352, 192}

\bibitem[\protect\citeauthoryear{{Maccarone}}{{Maccarone}}{2003}]{Maccarone2003}
{Maccarone} T.~J.,  2003, \mn@doi [\aap] {10.1051/0004-6361:20031146}, \href
  {https://ui.adsabs.harvard.edu/abs/2003A&A...409..697M} {409, 697}

\bibitem[\protect\citeauthoryear{{Matsuoka} et~al.,}{{Matsuoka}
  et~al.}{2009}]{Matsuoka2009}
{Matsuoka} M.,  et~al., 2009, \mn@doi [\pasj] {10.1093/pasj/61.5.999}, \href
  {https://ui.adsabs.harvard.edu/abs/2009PASJ...61..999M} {61, 999}

\bibitem[\protect\citeauthoryear{McClintock \& Remillard}{McClintock \&
  Remillard}{2009}]{McClintock2009}
McClintock J.~E.,  Remillard R.~A.,  2009, in Lewin W.,  van~der Klis M.,  eds,
  , Compact Stellar X-ray Sources.
Cambridge Univ. Press, pp 157--214, \mn@doi{10.1017/cbo9780511536281.005}

\bibitem[\protect\citeauthoryear{{McMullin}, {Waters}, {Schiebel}, {Young}  \&
  {Golap}}{{McMullin} et~al.}{2007}]{McMullin2007}
{McMullin} J.~P.,  {Waters} B.,  {Schiebel} D.,  {Young} W.,   {Golap} K.,
  2007, in {Shaw} R.~A.,  {Hill} F.,   {Bell} D.~J.,  eds,  Astron. Soc. Pac.
  Conf. Ser. Vol. 376, Astronomical Data Analysis Software and Systems XVI.
  p.~127

\bibitem[\protect\citeauthoryear{{Miller-Jones}, {Jonker}, {Dhawan}, {Brisken},
  {Rupen}, {Nelemans}  \& {Gallo}}{{Miller-Jones}
  et~al.}{2009}]{Miller-Jones2009}
{Miller-Jones} J.~C.~A.,  {Jonker} P.~G.,  {Dhawan} V.,  {Brisken} W.,  {Rupen}
  M.~P.,  {Nelemans} G.,   {Gallo} E.,  2009, \mn@doi [\apjl]
  {10.1088/0004-637X/706/2/L230}, \href
  {https://ui.adsabs.harvard.edu/abs/2009ApJ...706L.230M} {706, L230}

\bibitem[\protect\citeauthoryear{{Mirabel} \& {Rodr{\'\i}guez}}{{Mirabel} \&
  {Rodr{\'\i}guez}}{1994}]{Mirabel1994}
{Mirabel} I.~F.,  {Rodr{\'\i}guez} L.~F.,  1994, \mn@doi [\nat]
  {10.1038/371046a0}, \href
  {https://ui.adsabs.harvard.edu/abs/1994Natur.371...46M} {371, 46}

\bibitem[\protect\citeauthoryear{{Murphy}, {Mauch}, {Green}, {Hunstead},
  {Piestrzynska}, {Kels}  \& {Sztajer}}{{Murphy} et~al.}{2007}]{Murphy2007}
{Murphy} T.,  {Mauch} T.,  {Green} A.,  {Hunstead} R.~W.,  {Piestrzynska} B.,
  {Kels} A.~P.,   {Sztajer} P.,  2007, \mn@doi [\mnras]
  {10.1111/j.1365-2966.2007.12379.x}, \href
  {https://ui.adsabs.harvard.edu/abs/2007MNRAS.382..382M} {382, 382}

\bibitem[\protect\citeauthoryear{{Reid}, {McClintock}, {Narayan}, {Gou},
  {Remillard}  \& {Orosz}}{{Reid} et~al.}{2011}]{Reid2011}
{Reid} M.~J.,  {McClintock} J.~E.,  {Narayan} R.,  {Gou} L.,  {Remillard}
  R.~A.,   {Orosz} J.~A.,  2011, \mn@doi [\apj] {10.1088/0004-637X/742/2/83},
  \href {https://ui.adsabs.harvard.edu/abs/2011ApJ...742...83R} {742, 83}

\bibitem[\protect\citeauthoryear{{Reid} et~al.,}{{Reid}
  et~al.}{2014}]{Reid2014}
{Reid} M.~J.,  et~al., 2014, \mn@doi [\apj] {10.1088/0004-637X/783/2/130},
  \href {https://ui.adsabs.harvard.edu/abs/2014ApJ...783..130R} {783, 130}

\bibitem[\protect\citeauthoryear{{Reynolds}}{{Reynolds}}{1994}]{Reynolds:1994}
{Reynolds} J.,  1994, AT Technical Document AT/39.3/040

\bibitem[\protect\citeauthoryear{{Russell}, {Anderson}, {Miller-Jones},
  {Degenaar}, {Eijnden}, {Sivakoff}  \& {Tetarenko}}{{Russell}
  et~al.}{2019}]{Russell2019a}
{Russell} T.,  {Anderson} G.,  {Miller-Jones} J.,  {Degenaar} N.,  {Eijnden} J.
  v.~d.,  {Sivakoff} G.~R.,   {Tetarenko} A.,  2019, ATel, \href
  {https://ui.adsabs.harvard.edu/abs/2019ATel12456....1R} {12456, 1}

\bibitem[\protect\citeauthoryear{{Sault}, {Teuben}  \& {Wright}}{{Sault}
  et~al.}{1995}]{Sault1995}
{Sault} R.~J.,  {Teuben} P.~J.,   {Wright} M.~C.~H.,  1995, in {Shaw} R.~A.,
  {Payne} H.~E.,   {Hayes} J.~J.~E.,  eds,  Astron. Soc. Pac. Conf. Ser. Vol.
  77, Astronomical Data Analysis Software and Systems IV. p.~433 (\mn@eprint
  {arXiv} {astro-ph/0612759})

\bibitem[\protect\citeauthoryear{{Tetarenko} et~al.,}{{Tetarenko}
  et~al.}{2017}]{Tetarenko2017}
{Tetarenko} A.~J.,  et~al., 2017, \mn@doi [\mnras] {10.1093/mnras/stx1048},
  \href {https://ui.adsabs.harvard.edu/abs/2017MNRAS.469.3141T} {469, 3141}

\bibitem[\protect\citeauthoryear{{Tominaga} et~al.,}{{Tominaga}
  et~al.}{2020}]{Tominaga2020}
{Tominaga} M.,  et~al., 2020, \mn@doi [\apjl] {10.3847/2041-8213/abaaaa}, \href
  {https://ui.adsabs.harvard.edu/abs/2020ApJ...899L..20T} {899, L20}

\bibitem[\protect\citeauthoryear{{Tremou} et~al.,}{{Tremou}
  et~al.}{2020}]{Tremou2020}
{Tremou} E.,  et~al., 2020, \mn@doi [\mnras] {10.1093/mnrasl/slaa019}, \href
  {https://ui.adsabs.harvard.edu/abs/2020MNRAS.493L.132T} {493, L132}

\bibitem[\protect\citeauthoryear{{Vahdat Motlagh}, {Kalemci}  \&
  {Maccarone}}{{Vahdat Motlagh} et~al.}{2019}]{Vahdat2019}
{Vahdat Motlagh} A.,  {Kalemci} E.,   {Maccarone} T.~J.,  2019, \mn@doi
  [\mnras] {10.1093/mnras/stz569}, \href
  {https://ui.adsabs.harvard.edu/abs/2019MNRAS.485.2744V} {485, 2744}

\bibitem[\protect\citeauthoryear{{Wenger}, {Balser}, {Anderson}  \&
  {Bania}}{{Wenger} et~al.}{2018}]{Wenger2018}
{Wenger} T.~V.,  {Balser} D.~S.,  {Anderson} L.~D.,   {Bania} T.~M.,  2018,
  \mn@doi [\apj] {10.3847/1538-4357/aaaec8}, \href
  {https://ui.adsabs.harvard.edu/abs/2018ApJ...856...52W} {856, 52}

\bibitem[\protect\citeauthoryear{{Williams} et~al.,}{{Williams}
  et~al.}{2020}]{Williams2020}
{Williams} D.~R.~A.,  et~al., 2020, \mn@doi [\mnras] {10.1093/mnrasl/slz152},
  \href {https://ui.adsabs.harvard.edu/abs/2020MNRAS.491L..29W} {491, L29}

\bibitem[\protect\citeauthoryear{{Yatabe} et~al.,}{{Yatabe}
  et~al.}{2019}]{Yatabe2019}
{Yatabe} F.,  et~al., 2019, ATel, \href
  {https://ui.adsabs.harvard.edu/abs/2019ATel12425....1Y} {12425, 1}

\bibitem[\protect\citeauthoryear{{Zdziarski}, {Poutanen}, {Mikolajewska},
  {Gierlinski}, {Ebisawa}  \& {Johnson}}{{Zdziarski}
  et~al.}{1998}]{Zdziarski1998}
{Zdziarski} A.~A.,  {Poutanen} J.,  {Mikolajewska} J.,  {Gierlinski} M.,
  {Ebisawa} K.,   {Johnson} W.~N.,  1998, \mn@doi [\mnras]
  {10.1046/j.1365-8711.1998.02021.x}, \href
  {https://ui.adsabs.harvard.edu/abs/1998MNRAS.301..435Z} {301, 435}

\bibitem[\protect\citeauthoryear{{Zhang} et~al.,}{{Zhang}
  et~al.}{2020}]{LZhang2020}
{Zhang} L.,  et~al., 2020, arXiv e-prints, \href
  {https://ui.adsabs.harvard.edu/abs/2020arXiv200907749Z} {p. arXiv:2009.07749}

\makeatother
\end{thebibliography}
\bsp    
\label{lastpage}
\end{document}